\documentstyle[anta, twoside,12pt, epsfig]{article}

\begin{document}
\thispagestyle{plain}

\title{The Effect of aberration on polarization position angle of pulsars}
\author{D. Mitra$^1$ and J. H. Seiradakis$^2$} 

\address{$^1$Max-Planck-Institut f\"ur Radioastronomie, Auf dem H\"ugel 69, 53121 Bonn, Germany\\
$2$University of Thesseloniki,
Department of Physics,
Section of Astrophysics, Astronomy and Mechanics,
GR-54124 Thessaloniki, Greece}

\maketitle

\abstract{The linear polarization position angle (PPA) of pulsar
emission shows a wide variety of structures, which deviates from the classical
rotating vector model. In this study we investigate the effect on the PPA 
due to emission arising from a range of heights. We demonstrate that by appropriately
choosing the emission height we can reconstruct the PPA sweep across
the integrated pulse profiles.}

\section{Introduction}
Pulsars exhibit broad-band coherent radio emission ranging
from tens of MHz to tens of GHz and is highly polarized. 
It is usually observed that pulse widths at higher frequencies
are smaller than that at lower frequencies (see Mitra \& Rankin 2002 for
a recent review). This phenomenon is termed as radius to frequency
mapping (RFM) according to which emission at progressively higher frequencies
arise closer to the stellar surface. 
The polarization position angle (PPA) of the linear polarization
for several pulsars is seen to execute a smooth `S-shaped' curve
which has been interpreted as evidence for highly beamed emission
arising from dipolar magnetic field lines as per the 
rotating vector model (RVM) proposed by Radhakrishnan \& Cooke (1969).
Fitting the RVM model to the PPA traverse can be used to find
the geometrical angles of the neutron star namely $\alpha$ the angle 
between the rotation axis and the dipolar magnetic axis and $\beta$ the
angle between the rotation axis and the observers line-of-sight.
RVM and RFM together has been extensively used to find radio emission
heights in pulsars and the shape of the pulsar beam. Canonical values
for emission heights are found to be 50 times the stellar radius
and the emission beam is thought to be organized in the shape of 
nested cones of emission, with a central core emission.

Whilst RFM is more commonly observed in pulsars (barring a few 
exceptions) there are significant deviations from the RVM observed 
in PPA traverses in pulsars. In some cases across
the pulse the PPA traverse is seen to have 90$^{\circ}$ jumps and/or 
two orthogonal PPA tracks. Careful study has shown that each of the
PPA track are in agreement with the RVM (Gil \& Lyne 1994).
The other kind of deviation is in the form of wiggles and 
kinks (non-orthogonal departures) observed in
the PPA traverse. In this paper we will concentrate on the latter kind
of deviation observed and put forward a possible scenario which can 
explain such deviations.

\section{The BCW model and the RVM curve}
The RVM model has been further advanced by Blaskiewicz et al. (1991,
hereafter BCW) where they included first order special relativistic
effects into account. The model predicts that due to aberration and
retardation (A/R) effects the PPA traverse depends on the emission height $r_{em}$
and the pulsar period $P$ (in sec) as,
\begin{equation}
\psi=\psi_{0}+\arctan\left(\frac{\sin\alpha \sin(\phi-\phi_{0})-(6\pi/P)(r_{em}/v_c)\sin\xi}{\sin\xi
\cos\alpha+\sin\alpha \cos\xi\cos(\phi-\phi_{0})}\right).
\label{eq1}
\end{equation}
Here $\psi_{0}$ and $\phi_{0}$ are arbitrary position angle and longitude phase 
offsets and $\phi$ is the longitude of the pulse. The angle $\xi=\alpha+\beta$ and
$v_c$ is the velocity of light.
Note that the above equation reduces to the RVM for ${r_{em}}=0$.
It is immediately obvious from the above equation that
changes in ${r_{em}}$ across the pulse will result in distortion of the 
PPA traverse. Also if ${r_{em}}$ is fixed, then the distortions will be larger
for faster pulsars, as expected from the A/R effects.

However ${r_{em}}$ across the pulse is an unknown. It is thought that the 
central core emission arises much below the conal emission. In Figure 1 we show
an example of how the PPA curve will be distorted if emission from different
parts of the pulse originated from two distinctly different heights.
To refine the above concept of emission arising from different heights
we put forward the following model,

(a) We note a pulse profile can be decomposed as a set of Gaussians (Kramer 1994). 
Further each Gaussian can be attributed to either cone or core emission. These
core and cone can arise from different heights.

(b) For every i$^{th}$ longitude in the pulse, the resultant emission height ${r_{em}(i)}$
is an intensity (I) weighted quantity given as,
\begin{equation}
{r}_{em}(i) = \frac{\sum_{c=1}^{N} (I(i,c) \times r(c))}{\sum_{c=1}^{N} I(i,c)}
\label{eq2}
\end{equation}
Here the index $c$ refers to the Gaussian component and ${r(c)}$ corresponds
to the emission height attributed to that component (which correspond to height of the 
cone or core component).  In Figure 1 we show an example on how
the PPA traverse behaves when equation~\ref{eq2}  is applied. It should be noted that
the PPA traverse strongly depends on the parameters $\alpha$, $\beta$ and $ r_{em}$ 
which are highly correlated to each other.
Distortions of the PPA traverse
due to height dependent emission has also been investigated by Xu. et. al. (1997).

\begin{figure*}
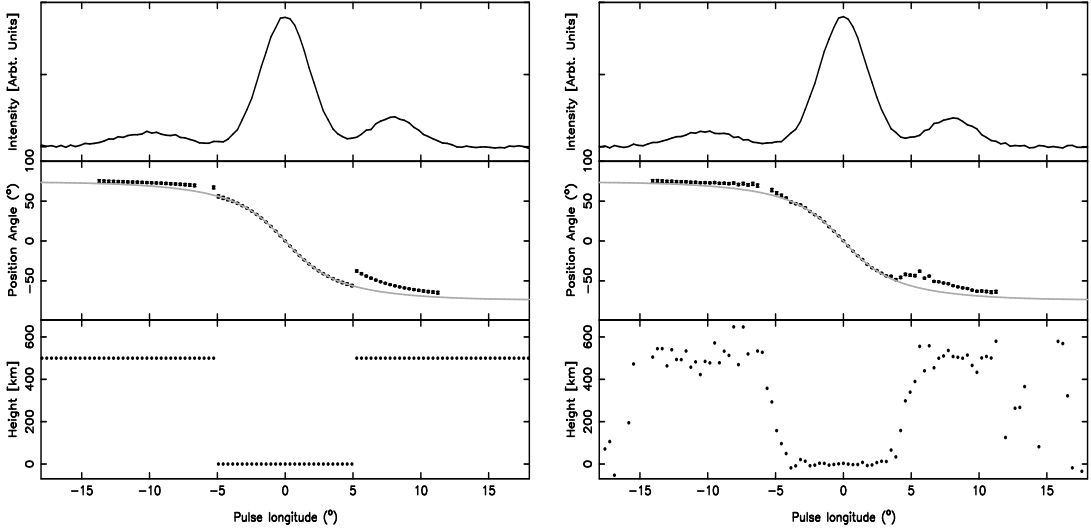

\begin{tabular}{@{}lr@{}}
{\mbox{\epsfig{file=g_fig1a.ps,width=7cm,height=7cm,angle=-90}}}&
{\mbox{\epsfig{file=g_fig1b.ps,width=7cm,height=7cm,angle=-90}}}\\
\end{tabular}
\caption{The above plot shows the total intensity (top panel) the PPA traverse (middle panel) 
and the emission height (bottom panel) for a SIMULATED pulsar of period 0.7 sec, $\alpha = 39^{\circ}$,
$\beta=2^{\circ}$ and can be adequately described by three Gaussian component 
giving rise to a central core and outer cone. The middle panel also shows the PPA traverse
corresponding to the RVM model in solid line where as the points correspond to the BCW model.
The left hand figure correspond to the case where two discrete emission height 
is considered. 
The right figure corresponds to the case where equation \ref{eq2} has been applied to obtain
the PPA traverse. Note that in both the case deviations in form of kinkiness appears in the 
PPA traverse (see text for further details)}
\label{fig1}
\end{figure*}

\section{Application of BCW model to pulsars}
Most of the time the distortions in the PPA traverse seem to occur below
the central core component in pulsars. While it is seen in a few normal 
pulsars (NP), it is more commonly seen in millisecond pulsars (MSP). Also in 
MSP's the PPA traverse is seen to be flat.

PSR J1022+1001 a millisecond pulsar with a period of 16 msec at 408 MHz is a classical 
example where the PPA traverse shows distortions below the central core component, 
as shown in
Figure 2 (left plot). The kinky nature of the PPA traverse seen below the central
core component. This pulsar can be adequately described by fitting three Gaussians, two
outer ones comprising of one cone and the inner core. On the right hand plot of Figure
2 we show simulation of the PPA traverse using equation~\ref{eq2} assuming that the central
core emission arises lower than the outer cone emission. This way we are able 
to generate the kinky feature in the PPA traverse. However we emphasize here
that given the simple model that is being used we are only able to match the
gross properties of the PPA traverse and do not claim to find the exact solution
for $\alpha$, $\beta$ and $r_{em}$. For this a more sophisticated model
needs to be considered. 

\begin{figure*}
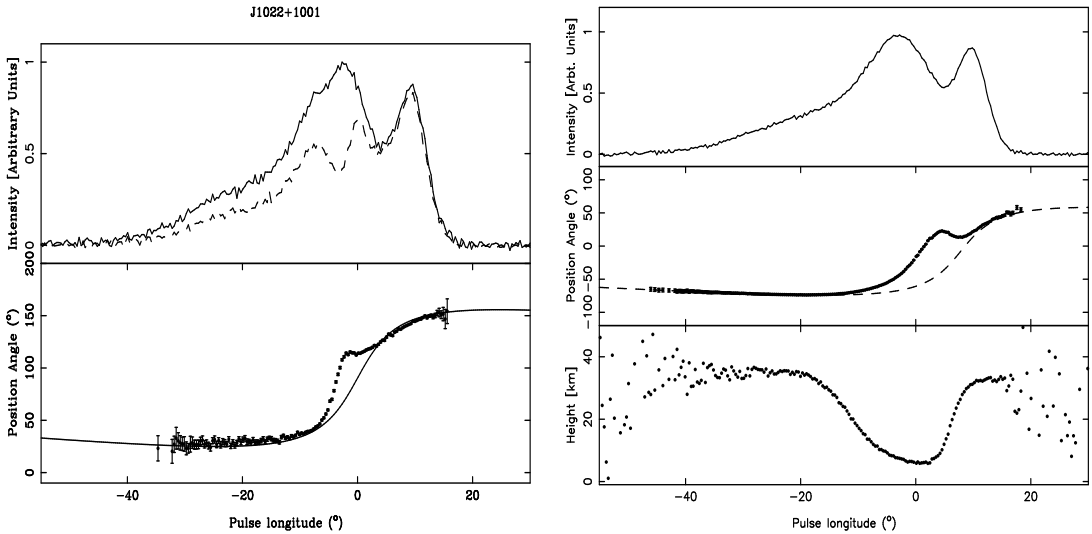

\begin{tabular}{@{}lr@{}}
{\mbox{\epsfig{file=g_fig2a.ps,width=7cm,height=7cm,angle=-90}}}&
{\mbox{\epsfig{file=g_fig2b.ps,width=7cm,height=7cm,angle=-90}}}\\
\end{tabular}
\caption{The top panel of the left plot shows the total intensity profile (solid line)
and linear polarization (dashed line) of PSR J1022+1001 at 410 MHz (data from Gould \& Lyne 1998
as obtained from the EPN archive maintained by MPIFR Bonn). The bottom panel shows the PPA
traverse and the dashed line shows the RVM curve described by $\alpha=3^{\circ}$ and $\beta=0.3^{\circ}$.
The right plot is a simulation and application of equation~\ref{eq2}. The meaning of the panels are
same as in Figure 1.}
\label{fig1}
\end{figure*}

\section{Conclusion}
Using a simple model we have been able to produce the kinkiness in the PPA traverse
towards the central region of the pulse profiles which is quite often observed.
As a consequence we are able to conclude the following:

(a) The central (or core) emission arises lower than the outer conal emission
confirming a suggestion put forward by several people (see for example Rankin 1993).

(b) The PPA traverse results from emission arising over a broad range of heights as 
has been recently suggested by Mitra \& Li (2003). 

Recently the central kinky feature seen in PSR J1022+1001 was interpreted as 
evidence for `return currents' in pulsars by Ramachandran \& Kramer (2003). Here 
we show that by considering height dependent emission regions similar kinky 
features can be obtained and it is not necessary to invoke return current
effects to explain this feature.

\section*{References}\noindent

\references
Blaskiewicz, M., Cordes, J.~M., \& Wasserman, I.\ 1991, ApJ, 370, 643 

Gil, J. A. \& Lyne, A. G. \ 1995, MNRAS, 276, L55

Gould, D.~M.~\& Lyne, A.~G.\ 1998, MNRAS, 301, 235

Mitra, D. \& Li, X. H., 2003, A\&A in press

Mitra, D.~\& Rankin, J.~M.\ 2002, ApJ, 577, 322

Kramer, M.\ 1994, A\&AS, 107, 527

Radhakrishnan, V.~\& Cooke, D.~J.\ 1969, ApJL, 3, 225

Ramachandran, R. \& Kramer, M., 2003, A\&A, 407, 1085

Rankin, J.~M.\ 1993, ApJ, 405, 285 

Xu, R. X., Qiao, G. J. \& Han, J. L., 1997, A\&A, 323, 395

\end{document}